\documentclass[pra,twocolumn,amsmath,amssymb,floatfix,reprint,footinbib,superscriptaddress,longbibliography,showkeys]{revtex4-2}%longbibliography
\usepackage[english]{babel}
\usepackage{picinpar,graphicx}
\usepackage{braket}
\usepackage{amsmath}
\usepackage{graphicx}
\usepackage{epstopdf}
\usepackage{dcolumn}
\usepackage{graphicx}
\usepackage{mathrsfs}
\usepackage{mdwlist}
\usepackage{subfigure}
\usepackage{booktabs}
\usepackage{amsmath}
\usepackage{dsfont}
\usepackage{amstext}
\usepackage{amssymb}
\usepackage{amsbsy}
\usepackage{bbm}
\usepackage{amsthm}
\usepackage{graphicx}
\usepackage{textcomp}
\usepackage{color}
\usepackage[colorlinks,citecolor=blue]{hyperref}
\usepackage{lipsum}
\definecolor{Dgreen}{RGB}{0, 100, 0}
\usepackage{url}
\usepackage[colorlinks]{hyperref}
\hypersetup{%
	plainpages=true,
	breaklinks=true,  %not default in dvips mode, so we must specify
	hypertexnames=false,  %not ideal, but needed when pagenums duplicate (`i' vs. `1')
	pageanchor=true,
	colorlinks=true,
	linkcolor={blue},
	citecolor={red},
	urlcolor={blue},
	anchorcolor={black}
}

\begin{document}

% --- TITLE AND AUTHOR INFORMATION ---
% Structure is critical here. Each author gets their own block.

\title{Relaxed parameter sensitivity for multiphoton quantum resonances}

% --- 第一组作者 (学生/研究人员) ---
\author{Hao-Lin Zhong}
\author{Ke-Xiong Yan}
\author{Yi-Ming Yu}
\author{Shao-Wei~Xu}
% --- 上述作者共享的单位地址 ---
\affiliation{Fujian Key Laboratory of Quantum Information and Quantum Optics, College of Physics and Information Engineering, Fuzhou University, Fuzhou 350108, China}

\author{Zhi-Cheng Shi}
\email{szc20147@163.com}
\affiliation{Fujian Key Laboratory of Quantum Information and Quantum Optics, College of Physics and Information Engineering, Fuzhou University, Fuzhou 350108, China}

% --- 第二组作者 (通讯作者) ---
\author{Ye-Hong Chen}
\email{yehong.chen@fzu.edu.cn}
% --- 上述通讯作者共享的单位地址 ---
\affiliation{Fujian Key Laboratory of Quantum Information and Quantum Optics, College of Physics and Information Engineering, Fuzhou University, Fuzhou 350108, China}
\affiliation{Institute of Quantum Science and Technology, Yanbian University, Yanji 133002, China}
\affiliation{Quantum Information Physics Theory Research Team, Center for Quantum Computing, RIKEN, Wako-shi, Saitama 351-0198, Japan}

\author{Yan Xia}
\email{xia-208@163.com}
\affiliation{Fujian Key Laboratory of Quantum Information and Quantum Optics, College of Physics and Information Engineering, Fuzhou University, Fuzhou 350108, China}
\affiliation{Institute of Quantum Science and Technology, Yanbian University, Yanji 133002, China}

\date{\today}

% --- ABSTRACT ---
% The abstract environment MUST come before the \maketitle command.
\begin{abstract}
Multiphoton resonances demonstrate the physical significance of counter-rotating wave terms in light-matter interactions. These resonances, however, are sensitive to detuning errors, making the phenomena challenging to experimentally observe. In this manuscript, we introduce an optimization strategy to address this problem. By using an optimized parameter segmented sequence (OPSS), the robustness against detuning errors of the high-order quantum state transfers can be substantially improved. We prove the versatility of our strategy against frequency detunings by demonstrating the evolution of two specific models. In both cases, the parameter window for maintaining a high state-transfer fidelity is substantially expanded. We further analyze the output photon flux of the optimized system and, taking the three-photon resonance as an example, demonstrate that the system remains capable of generating a stable output photon flux even in the presence of detuning errors. 
\end{abstract}

% --- MAKETITLE ---
% This command generates the formatted title, author list, and abstract.
% It MUST be placed after the \abstract block.
\maketitle
\section{Introduction}
\label{sec:intro}

The research of nonlinear optics is founded on the principle that a medium can respond nonlinearly to intense incident light. This principle develops a host of phenomena, which have not only found critical technological applications \cite{Lindsay1975, Kielich1981, Shen1984, Boyd2008}, but have also been demonstrated in a vast range of nonlinear media \cite{Liu2022, Dudley2006, Acharyya2021, Kumar2021}. The profound implications of nonlinearity have inspired investigations into analogous effects in other wave-based systems, with prominent examples found in nonlinear acoustics \cite{2024}, atom optics \cite{Rolston2002} and nonlinear Josephson plasma waves \cite{Savelev2006}. 

Analogs of such nonlinear optical effects are readily accessible in cavity quantum electrodynamics systems \cite{Blais2021}, where one or more resonator modes are coupled to a two-level atom (or an artificial atom). Just as conventional nonlinear optical effects typically require high light intensity to manifest clearly, the higher-order processes also demand a strong light-matter coupling strength \cite{Niemczyk2010}. This requirement is met in the ultrastrong coupling regime \cite{Niemczyk2010,Fedorov2010,FornDaz2010, FornDaz2016}, where the coupling strength becomes comparable to the resonant frequencies of the system \cite{FriskKockum2019, FornDaz2019}. Such significant coupling strength renders the counter-rotating terms non-negligible, thereby opening additional transition channels for higher-order resonant processes mediated by virtual intermediate states \cite{FornDaz2010}. These high-order processes allow efficient energy exchange between the qubit and the resonator even in the large detuning regime where single-photon transitions are suppressed \cite{Garziano2015}. The existence of these excitation-number-nonconserving pathways has excited numerous theoretical investigations, such as multi-photon oscillation \cite{Garziano2015} and multi-atom excitation \cite{Ma2015, Garziano2016}.  However, achieving such a regime is experimentally demanding, often requiring extreme system parameters or complex fabrication techniques to sufficiently enhance the coupling strength \cite{Mueller2025, FriskKockum2019}. 

In this manuscript, we investigate multi-photon resonances in the large-detuning regime with relatively weak coupling. We primarily focus on the three-photon Rabi resonance \cite{Ma2015}, involving the exchange of a qubit excitation with three photons. Furthermore, we extend our scope to the Casimir-Rabi resonance \cite{Yan2025}, where two photons are converted into three phonons. While these phenomena are of fundamental interest, the higher-order resonances in this regime are inherently sensitive to experimental errors. This is because the high-order processes rely on weak effective couplings, which makes them extremely fragile against any fluctuations in system parameters.

To overcome this fragility, we propose an optimized parameter segmented sequence (OPSS), i.e., a method inspired by composite pulse techniques \cite{Kyoseva2019, Torosov2019}. Originating from nuclear magnetic resonance \cite{Freeman1998,Levitt2007}, composite pulses have become a standard tool for robust coherent control in diverse systems, including trapped ions \cite{HAFFNER2008, Monz2009, Shappert2013, Timoney2008, Gulde2003, Zarantonello2019,Mount2015}, atoms \cite{Rakreungdet2009, Demeter2016}, and quantum dots \cite{Wang2012,Eng2015, Genov2017, Wang2014,Zhang2017,Kestner2013, Hickman2013}.

\begin{figure}[tbp]
    \centering
    \includegraphics[width=0.9\columnwidth]{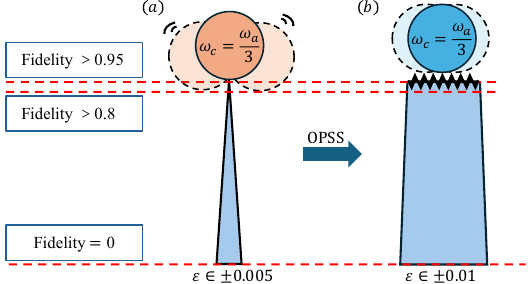}
    \caption{Schematic comparison of the robustness against detuning frequency errors for the three-photon resonance. 
    (a) The unoptimized case. The resonance is observable only under stringent parameter conditions. The fidelity decreases to zero at a detuning error of $\varepsilon = \pm 0.5\%$. Specifically, the fidelity reduces to $0.9$ with a minute deviation of $\epsilon = 0.05\%$. 
    (b) The optimized case (OPSS). The system maintains the observability of the resonance over a significantly broader error range, with fidelity degradation occurring only beyond $\varepsilon = \pm 1\%$. Within this interval, the rapid oscillations maintain the fidelity in the range $[0.8, 0.9]$, ensuring the experimental visibility of the resonance.}
\label{fig:introduce} %%% It is good practice to keep a label
\end{figure}

Analogous to composite pulses that compensate for errors via varying phases or detunings \cite{Torosov2011, Kyoseva2019}, we propose a frequency-domain strategy. Instead of a static configuration, we construct an OPSS where cavity detunings serve as dynamic control parameters optimized to render the system less sensitive to parameter fluctuations. Since analytical optimization is intractable, we employ Differential Evolution (DE)~\cite{Bilal2020, Ahmad2022} combined with GRAPE~\cite{Khaneja2005, Machnes2011} to numerically determine the ideal OPSS. Our results show that the proposed OPSS method dramatically improves the tolerance to cavity frequency errors. For the standard three-photon resonance, a mere $0.5\%$ error triggers a serious drop in fidelity from ${\sim}92\%$ to ${\sim}1\%$, which leads to the disappearance of the resonance signal. In contrast, our method sustains the fidelity at a high level via rapid oscillatory dynamics. Overall, this enables the resonance phenomenon to be observed within a significantly broader error range Fig.\ref{fig:introduce}.

Furthermore, to demonstrate the practical viability of OPSS, we extend our analysis to include environmental dissipation. By analyzing the cavity output photon flux, we observe that the optimized system yields a stable signal even in error-prone regions where the unoptimized counterpart fails to produce detectable emission.

The paper is organized as follows. Section~\ref{sec:theory} outlines the theoretical models and physical conditions for the three-photon and Casimir-Rabi resonances. Section~\ref{sec:method} details the optimization strategy based on OPSS. Section~\ref{sec:imply} analyzes the results, demonstrating how the optimized sequences render the system less sensitive to detuning errors. Finally, Section~\ref{sec:conclusion} concludes the work.
% =======================================================================
 %%%%%%%%%%%%%%%%%%%%%%%%%%%%%%%%%%%%%%%%%%%%%%%%%%%%%%%%%%%%%%%%%%%%%%%%%%%%%%%%%
% --- SECTION II A: REVISED VERSION ---
%%%%%%%%%%%%%%%%%%%%%%%%%%%%%%%%%%%%%%%%%%%%%%%%%%%%%%%%%%%%%%%%%%%%%%%%%%%%%%%%%

\section{THEORETICAL MODELS}
\label{sec:theory}

\subsection{Three-Photon Resonance}
\label{sec:model_3p}

We begin by considering the quantum Rabi model ($\hbar=1$), which describes the interaction between a two-level system (with ground state $|g\rangle$ and excited state $|e\rangle$) and a single-mode cavity resonator \cite{Ma2015}. The Hamiltonian is given by
\begin{equation}
    H_R = \frac{\omega_a}{2}\sigma_z + \omega_c a^\dagger a + \lambda(a + a^\dagger)\sigma_x.
    \label{eq:Rabi_Hamiltonian}
\end{equation}
Here, $\omega_a$ and $\omega_c$ are the frequencies of the qubit and the cavity mode, respectively; $a$ ($a^\dagger$) is the annihilation (creation) operator for the cavity mode, and $\lambda$ is the coupling strength. The Pauli matrices are defined as $\sigma_z = \ket{e}\bra{e} - \ket{g}\bra{g}$ and $\sigma_x = \ket{e}\bra{g} + \ket{g}\bra{e}$. 

\begin{figure}[tbp]
    \centering
    \includegraphics[width=0.9\columnwidth]{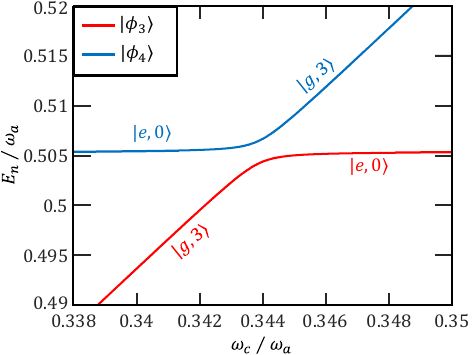}
    \caption{Eigenvalues of the third ($\ket{\phi_3}$) and fourth ($\ket{\phi_4}$) eigenstates are plotted as a function of the frequency ratio $\omega_c/\omega_a$. An avoided crossing is observed at $\omega_c/\omega_a \approx 0.334$ between the bare states $\ket{e,0}$ and $\ket{g,3}$. The simulation uses a coupling strength of $\lambda = 0.06\,\omega_a$.}
\label{fig:energy_spectrum} %%% It is good practice to keep a label
\end{figure}

In the weak-coupling regime ($\lambda \ll \omega_a, \omega_c$), an avoided crossing emerges between bare states $\ket{e,0}$ and $\ket{g,3}$~\cite{Yan2024}, corresponding to eigenstates $\ket{\phi_3}$ and $\ket{\phi_4}$ in Fig.~\ref{fig:energy_spectrum}. The sensitivity of this gap to system parameters poses a primary challenge for high-fidelity operations \cite{Yan2024}. Near the resonance $\omega_c \approx \omega_a/3$, the dynamics are described by an effective Hamiltonian in the subspace spanned by $\{\ket{e,0}, \ket{g,3}\}$, derived via the time-averaging method~\cite{Shao2017} (see Appendix~\ref{app.A}).

\begin{align}
    H_{\text{eff}} &= \frac{3\lambda^2}{2\omega_a} \ket{e,0}\bra{e,0} \nonumber \\ 
                   &\quad + \left(- \frac{\lambda^2}{\omega_c} - \frac{9\lambda^2}{2\omega_a}\right)\ket{g,3}\bra{g,3} \nonumber \\ 
                   &\quad - \frac{9\sqrt{6}\lambda^3}{4\omega_a^2} \left( \ket{e,0}\bra{g,3} + \ket{g,3}\bra{e,0} \right) \nonumber \\ % 此处添加 \nonumber 去掉这一行的编号
                   &= 
    \begingroup % 使用 group 确保这个间距设置只影响当前这个矩阵
    \setlength{\arraycolsep}{10pt} % 【关键】这里设置列间距，数字越大越宽
    \begin{pmatrix}
        \dfrac{3\lambda^2}{2\omega_a} & - \dfrac{9\sqrt{6}\lambda^3}{4\omega_a^2} \\[5ex] % 稍微增加了行高以适应分数
        - \dfrac{9\sqrt{6}\lambda^3}{4\omega_a^2} & - \dfrac{\lambda^2}{\omega_c} - \dfrac{9\lambda^2}{2\omega_a}
    \end{pmatrix} 
    \endgroup .
    \label{eq:matrix}
\end{align}

The matrix form allows us to identify the effective detuning $\Delta$ and the Rabi frequency $\Omega_{\text{eff}}$ as
\begin{equation}
    \Delta = \frac{\lambda^2}{\omega_c} + \frac{6\lambda^2}{\omega_a} \quad \text{and} \quad
    \Omega_{\text{eff}} = - \frac{9\sqrt{6}\lambda^3}{4\omega_a^2}.
\end{equation}
The 2 $\times$ 2 matrix in Eq.~\ref{eq:matrix} fully describes the dynamics of the coupled two-level system, where the diagonal terms represent the AC Stark shifts of the states $\{\ket{e,0}, \ket{g,3}\}$ and the off-diagonal terms represent the effective coupling strength between them.

\subsection{Casimir-Rabi Resonance in a Nonlinear Optomechanical System}
\label{sec:model_cr}

As a second platform to test the generality of our method, we consider a nonlinear optomechanical system, analogous to a cavity whose resonant frequency varies with the displacement of a mechanical oscillator \cite{Law1995}. The Hamiltonian of Casimir-Rabi resonance is given by
\begin{equation}
    H_{op} = \omega_c a^\dagger a + \omega_m b^\dagger b + g(a^\dagger + a)^2(b^\dagger + b).
    \label{eq:optomech_hamiltonian}
\end{equation}
Here, $b$ ($b^\dagger$) is the annihilation (creation) operator for the mechanical mode of frequency $\omega_m$. The parameter $g$ represents the nonlinear coupling strength. The notation for the bare states is $\ket{n_c, n_m}$, representing $n_c$ photons and $n_m$ phonons.

\begin{figure}[tbp]
    \centering
    \includegraphics[width=0.9\columnwidth]{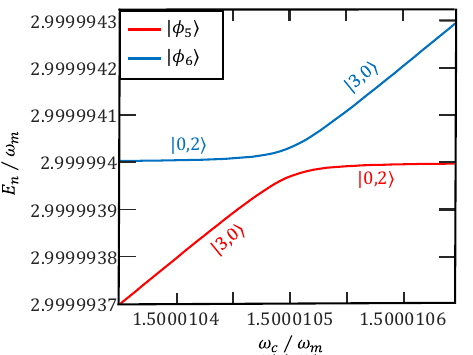}
    \caption{Eigenvalues of the fifth ($\ket{\phi_5}$) and sixth ($\ket{\phi_6}$) eigenstates are plotted as a function of the frequency ratio $\omega_c/\omega_m$. An avoided crossing is observed at $\omega_c/\omega_m \approx 1.5$ between the bare states $\ket{2,0}$ and $\ket{0,3}$. The minimum energy splitting is found at $\omega_c/\omega_m \approx 1.5000105$. The simulation uses a weak coupling strength of $g = 0.001\,\omega_m$.}
    \label{fig:kaxi_E} 
\end{figure}

Similar to the quantum Rabi model, this Hamiltonian illustrates a high-order multiphoton-multiphonon exchanging process when satisfying certain resonant conditions \cite{Macr2018, Yan2025}. The process of interest is visualized in the energy spectrum in Fig.~\ref{fig:kaxi_E}, where an avoided crossing is observed between the fifth ($\ket{\phi_5}$) and sixth ($\ket{\phi_6}$) eigenstates, corresponding to a coupling between the bare states $\ket{2,0}$ and $\ket{0,3}$ \cite{Yan2025}.

Projecting $H_s$ onto the basis $\{ \ket{2,0}, \ket{0,3} \}$ yields the effective Hamiltonian describing the subspace dynamics (see Appendix~\ref{app.A} for derivation):
\begin{align}
    H_{\text{eff}} &= \left(2\omega_c - \frac{27g^2}{\omega_m}\right) \ket{2,0}\bra{2,0} \nonumber \\
                   &\quad + \left(3\omega_m - \frac{6g^2}{\omega_m}\right) \ket{0,3}\bra{0,3} \nonumber \\
                   &\quad + \frac{18\sqrt{3}g^3}{\omega_m^2} \left( \ket{2,0}\bra{0,3} + \ket{0,3}\bra{2,0} \right) \nonumber \\ % 修改：去掉了 label，加上了 nonumber
                   &= 
    \begingroup % 开始局部设置
    \setlength{\arraycolsep}{10pt} % 【修改】设置列宽，数值越大越宽
    \begin{pmatrix}
        2\omega_c - \dfrac{27g^2}{\omega_m} & \dfrac{18\sqrt{3}g^3}{\omega_m^2} \\[3ex]
        \dfrac{18\sqrt{3}g^3}{\omega_m^2} & 3\omega_m - \dfrac{6g^2}{\omega_m}
    \end{pmatrix} 
    \endgroup . \label{eq:casimir_matrix_multiline}
\end{align}
Applying the similar procedure to the matrix of the Casimir-Rabi model, we can identify the effective detuning $\Delta$ and the effective Rabi frequency $\Omega_{\text{eff}}$:
\begin{equation}
    \Delta =3\omega_m - 2\omega_c + \frac{21g^2}{\omega_m} \quad\text{and}\quad
    \Omega_{\text{eff}} = \frac{18\sqrt{3}g^3}{\omega_m^2}.
\end{equation}

\subsection{Physical Origin of Detuning Sensitivity}
To elucidate the physical origin of this sensitivity, we analyze the population dynamics using the standard Rabi solution for an effective two-level system ~\cite{Ivanov2022}. The transition probability is 
\begin{equation}
    p = \frac{\Omega_{\text{eff}}^2}{\Omega_{\text{eff}}^2 + \Delta^2} \sin^2\left(\frac{\sqrt{\Omega_{\text{eff}}^2 + \Delta^2}}{2} t\right),
    \label{eq:Rabi_detuned}
\end{equation}
The ideal resonant conditions allows for a complete state flip ($p=1$). However, this equation also reveals the effect introduced by the detuning error $\epsilon$. When a small detuning error $\varepsilon$ is introduced (i.e., $\omega_a \rightarrow \omega_a(1+\varepsilon)$), it induces an error in $\Delta$, resulting in incomplete population transfer $p \neq 1$ ~\cite{Torosov2011}.

Eq.~\ref{eq:Rabi_detuned} indicates that the Casimir-Rabi resonance is far more fragile than the three-photon case. Since $\Omega_{\text{eff}}$ scales cubically with the base coupling strength ($\Omega_{\text{eff}} \propto g^3, \lambda^3$), the Casimir-Rabi resonance possesses an extremely weak optomechanical coupling ($g/\omega_m \approx 0.001$), which is orders of magnitude smaller than the three-photon coupling ($\lambda/\omega_a \approx 0.06$). Therefore, even a minute detuning error $\varepsilon$ causes the effective detuning term $\Delta$ to become dominant compared to the small $\Omega_{\text{eff}}$ in the Casimir system.

\section{OPTIMIZATION METHODOLOGY}
\label{sec:method}

To counteract the pronounced parameter sensitivity identified in Sec.~\ref{sec:theory}, we investigate and implement a numerically robust control strategy. Our approach departs from methods like composite pulses \cite{Ivanov2022}. Rather than designing controls for an effective Hamiltonian model, we directly optimize the parameters within the reduced Hamiltonian Eq.~\eqref{eq:Rabi_Hamiltonian} or \eqref{eq:optomech_hamiltonian}. Our approach replaces the single, static cavity frequency $\omega_c$ with a piecewise-constant sequence of optimized $\omega_c$ values. Critically, both the frequency values in this sequence and its total duration are numerically discovered by a hybrid DE + GRAPE optimization algorithm. This direct numerical optimization allows for the discovery of highly effective control protocols within a broader Hilbert subspace to achieve less-sensitive quantum state transfer in the studied models.

Our control scheme is defined by transforming the cavity frequency and the total evolution time into a set of optimizable parameters. The total evolution time $T$ is divided into $N$ segments of equal duration. During each segment $k$ (where $k=1, \dots, N$), the cavity frequency is held at a constant value $\omega_{c,k}$, making the Hamiltonian piecewise-static. The complete set of control parameters to be optimized is therefore the vector $\{\omega_{c,1}, \dots, \omega_{c,N}, T\}$, containing $N+1$ variables.
\begin{figure}[tbp]
    \centering
    \includegraphics[width=\columnwidth]{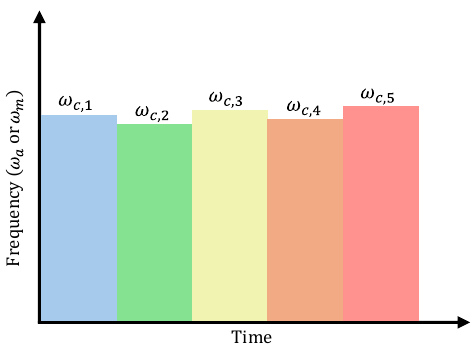}
    \caption{Schematic illustration of a set of optimizable parameters with different frequencies.}
    \label{fig:pulse} 
\end{figure} 

To address the high computational cost of simulating the reduced Hilbert space, we employ a numerical scheme designed to simplify the optimization process. Our approach relies on the direct linear relationship between the physical control frequencies and the effective detuning $\Delta$. This implies that optimizing the physical frequency sequence is equivalent to optimizing the $\Delta$ within the effective two-level framework. This strategy allows us to search for OPSS for the original physical model with significantly reduced computational cost. 

We scanned the fidelity profile of the target state as a function of $\varepsilon$. The optimization range was then defined by the boundaries at which the baseline fidelity drops to a negligible value. For the three-photon resonance, the interval is set to $[-0.5\%, 0.5\%]$. For the more fragile Casimir-Rabi resonance, a much narrower interval of $[-5 \times 10^{-6}\%, 5\times 10^{-6}\%]$ is chosen.

For the cost function evaluation, we uniformly sample $M$ points within the defined $\varepsilon$ range. The performance of each OPSS candidate is quantified by a composite cost function $C$, the minimization of which drives the optimization process (as illustrated in Fig.~\ref{fig:placeholder}). This function is defined as a weighted sum of three distinct components:
\begin{figure}
    \centering
    \includegraphics[width=\linewidth]{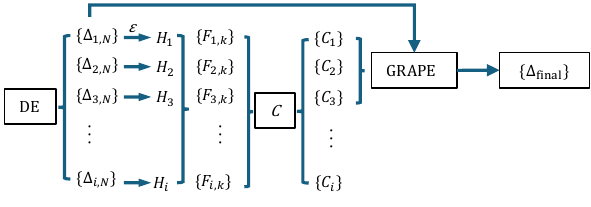}
    \caption{Schematic of the optimization workflow. The DE algorithm generates a population of $P=100$ candidate sequences $\{\Delta_{i,N}\}$ ($i=1,\dots,P$), where $N$ denotes the segment number. Each sequence constructs a Hamiltonian $H_i$ subject to detuning errors $\varepsilon$. The resulting fidelities $\{F_{i,k}\}$ are calculated across $M=51$ error sample points ($k$) to evaluate the cost score $C_i$. Finally, the top-performing candidates are refined by GRAPE to yield the final optimized sequence $\{\Delta_{\text{final}}\}$. (Parameters: $w_b=1, w_f=500, w_r=5$; max iterations: 1500 for DE, 3000 for GRAPE).}
    \label{fig:placeholder}
\end{figure}
\begin{equation}
    C = w_b C_{\text{barrier}} + w_f C_{\text{floor}} + w_r C_{\text{robust}}.
    \label{eq:cost_function}
\end{equation}
Here, $w_b$, $w_f$, and $w_r$ are weighting coefficients. Each component is calculated based on the set of fidelities $\{F_k\} (k=1, 2, \dots, M)$, where $F_k$ denotes the fidelity evaluated at the $k$-th sample point.

The first term $C_{\text{barrier}}$ acts as the primary driving force for maximizing performance. By imposing a logarithmic penalty, it pushes every sampling point toward the theoretical ideal value:
\[ C_{\text{barrier}} = \sum_{k=1}^{M} \log(1 - F_k). \]

The second term $C_{\text{floor}}$ establishes a minimum performance baseline. By selectively penalizing fidelities that fall below a critical threshold $F_{\text{target}}$, it compels the optimizer to prioritize the rectification of worst-case regions, thereby preventing significant degradations in the fidelity profile:
\[ C_{\text{floor}} = \sum_{k=1}^{M} \left[ \max(0, F_{\text{target}} - F_k) \right]^2. \]
The final term $C_{\text{robust}}$ directly quantifies the stability of the operation. Minimizing this standard deviation term flattens the fidelity profile, thereby enhancing the robustness of the pulse sequence against detuning errors:
\[ C_{\text{robust}} = \sqrt{\frac{1}{M} \sum_{k=1}^{M} (F_k - \bar{F})^2}, \]
Here, $\bar{F}$ represents the mean fidelity calculated over the $M$ sample points:
\[ \bar{F} = \frac{1}{M} \sum_{k=1}^{M} F_k. \]

Given the system's hypersensitivity and the resulting non-convex control landscape, we employ a two-stage hybrid strategy to ensure robust convergence. In the first stage, we utilize the DE algorithm for a global search. As a population-based stochastic evolutionary algorithm, DE does not rely on gradient information but instead evolves a large number of candidates simultaneously \cite{Storn1997, Das2011}. This characteristic enables DE to effectively navigate the vast parameter space and identify promising regions for the global optima, thereby avoiding local traps, while a callback function tracks and maintains an elite pool of the best candidates found. Subsequently, these top candidates serve as high-quality seeds for the second stage, where the GRAPE framework is tasked with the final high-precision refinement. GRAPE is a gradient-based optimal control algorithm known for its rapid convergence when initialized near a solution. We employ the L-BFGS-B algorithm~\cite{Byrd1995, Zhu1997} as the optimization engine. This stage acts as a final sprint, rapidly converging the pre-screened candidates to the optimal solution, after which we select the result with the lowest cost function value.

In this work, we specifically investigate the performance of the OPSS for segment numbers $N=3, 5$, and $7$. We find that increasing $N$ beyond $7$ yields diminishing returns in reducing sensitivity; the marginal improvement over the $N=7$ case does not justify the significantly higher computational cost.

\section{NUMERICAL IMPLEMENTATION}
\label{sec:imply}
\subsection{Parameter optimization in three-photon resonance}
\begin{figure}[tbp]
    \centering
    \includegraphics[width=\columnwidth]{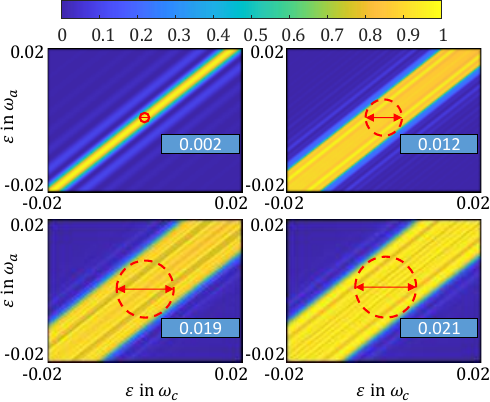} 
    \caption{Fidelity landscape under detuning error $\varepsilon$ applied to both $\omega_a$ and $\omega_c$. Each panel shows the performance for a different number of segments: (a) $N=1$, (b) $N=3$, (c) $N=5$, and (d) $N=7$. Red dashed circles mark the robust high-fidelity radius, whose expansion with increasing $N$ visualizes the enlarged error tolerance and reduced system sensitivity.}
    \label{fig:three_error}
\end{figure}

We begin our numerical analysis with the three-photon resonance model, establishing the unoptimized, fixed single-parameter case ($N = 1$) as the performance benchmark. To clearly demonstrate the performance enhancement achieved by our scheme, we scan the fidelity variation of the target state when applying detuning errors to both qubit frequency $\omega_a$ and cavity frequency $\omega_c$. Specifically, the perturbed frequencies are modeled as $\omega_a' = \omega_a(1+\varepsilon)$ and $\omega_c' = \omega_c(1+\varepsilon)$. The results are presented in Fig.~\ref{fig:three_error}, with each panel corresponding to a different number of parameter segments ($N=1, 3, 5, 7$).

% --- Figure 2 ---

For the benchmark case shown in Fig.~\ref{fig:three_error}(a), the high-fidelity region (yellow) is confined to an extremely narrow diagonal line. This indicates that high fidelity only exists when the parameters satisfy the strict resonance condition $\omega_a \approx 3\omega_c$. Any deviation from this stringent condition causes a severe drop. In contrast, the implementation of the OPSS strategy successfully reshapes the line into an exceptionally broad high-fidelity plateau. We set red dashed circles to indicate the effective high-fidelity region where $F > 0.8$. As the sequence $N$ increases, we observe a significant expansion of the radius, which directly corresponds to a broadened range of permissible errors. This achievement demonstrates that the sensitivity to detuning errors is effectively suppressed by the OPSS strategy.

To visually quantify the performance gains offered by different segment numbers $N$, we further investigate fidelity statistics under different levels of detuning error. To demonstrate the impact of individual parameter drifts, we fix the error on one parameter (e.g., $\omega_a$) at $\varepsilon=0$ while applying detuning errors to the other, then evaluate the performance within intervals centered around four distinct magnitudes: the ideal case ($\varepsilon=0$), a tiny error regime ($\varepsilon=0.1\%$), a moderate error regime ($\varepsilon=\pm 0.5\%$), and a large error regime ($\varepsilon=\pm 1\%$). The fidelity statistics are derived from a $\pm 10\%$ variation range surrounding each target value because the OPSS sustains the high performance via high-frequency oscillatory behavior.

\begin{figure*}[t]
    \centering
    \begin{minipage}[b]{0.49\textwidth}
        \centering
        \includegraphics[width=\linewidth]{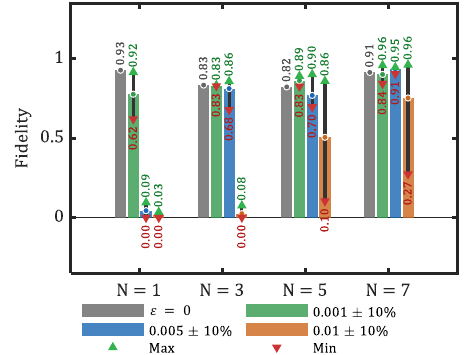} 
    \end{minipage}
    \hfill 
    \begin{minipage}[b]{0.49\textwidth}
        \centering
        \includegraphics[width=\linewidth]{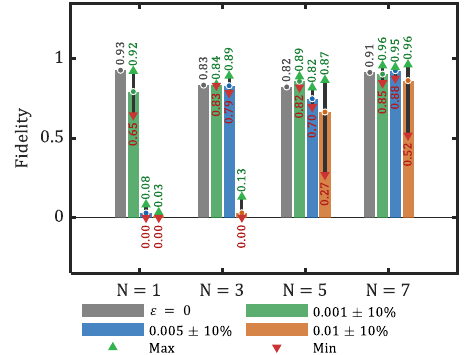}
    \end{minipage}
    \caption{Performance and robustness analysis. (a) Statistical performance analysis of $\varepsilon$ on $\omega_c$ (with $\varepsilon$ on $\omega_a=0$). (b) Statistical robustness analysis of $\varepsilon$ on $\omega_a$ (with $\varepsilon$ on $\omega_c=0$). The height of each bar represents the mean fidelity, while the green upward triangles ($\blacktriangle$) and red downward triangles ($\blacktriangledown$) indicate the maximum and minimum fidelities within the defined ranges.}
    \label{fig:combined_analysis}
\end{figure*}

The statistical results in Fig.~\ref{fig:combined_analysis} clearly demonstrate the effectiveness of the OPSS approach. For the benchmark case ($N=1$), the system is highly sensitive to parameter deviations. According to the green bar charts in Fig.~\ref{fig:combined_analysis}, even at a small error of $\varepsilon = 0.1\%$, the mean fidelity drops significantly to the $0.62 \sim 0.65$ range. As the error increases to $\varepsilon = 0.5\%$ (see the blue bar charts in Fig.~\ref{fig:combined_analysis}), the fidelity collapses to negligible levels ($F < 0.1$) for both $\omega_a$ and $\omega_c$ errors.

As $N$ increases, the system maintains high fidelities over a wider error range. Both the $N=3$ and $N=5$ sequences successfully protect the system against small errors, maintaining fidelities above $0.8$ at $\varepsilon = 0.1\%$ as shown by the green bar charts in Fig.~\ref{fig:combined_analysis}. However, the $N=3$ sequence begins to degrade when the error reaches $\varepsilon = 0.5\%$. The $N=5$ sequence extends this protection further but eventually declines at larger error magnitudes, particularly for $\omega_c$ errors at $\varepsilon = 1\%$, where the mean fidelity drops to around $0.1$.

The $N=7$ sequence provides the most comprehensive protection. It remains stable at $\varepsilon = 0.1\%$ and maintains high mean fidelities at $\varepsilon = 0.5\%$ ($0.88$ for $\omega_a$ and $0.91$ for $\omega_c$). At the large error magnitude of $\varepsilon = \pm 1\%$, which is demonstrated as the orange bar charts in Fig.~\ref{fig:combined_analysis}, although the average fidelity decreases, the maximum fidelity remains very high (${\sim}0.96$) in both cases. This confirms that the OPSS strategy ensures reliable state transfer is still achievable even when the system faces significant frequency deviations, without sacrificing the stability of one parameter for the other.

\subsection{Parameter optimization in the Casimir-Rabi Resonance}
To further test the versatility of our optimization, we apply it to the Casimir-Rabi resonance, a system with a distinct physical mechanism as discussed in Sec.~\ref{sec:theory}. As established previously, this model represents a more stringent benchmark due to its extremely weak nonlinear coupling strength ($g/\omega_m = 0.001$), which renders it orders of magnitude more sensitive to parameter errors than the three-photon model. Our goal is to implement a robust state transfer between the initial state $\ket{2,0}$ and the target state $\ket{0,3}$.

\begin{figure}[tbp]
    \centering
    \includegraphics[width=1\columnwidth]{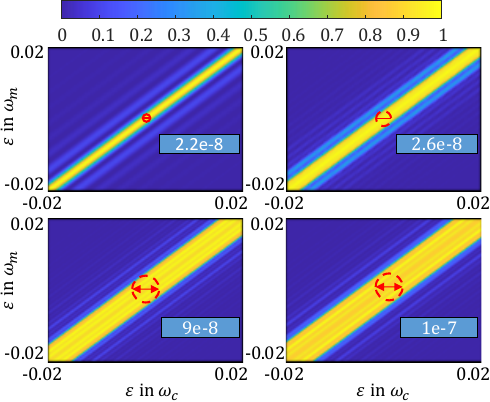} 
    \caption{Fidelity landscape under detuning error $\varepsilon$ in both $\omega_m$ and $\omega_c$. Each panel corresponds to a different number of pulse segments: (a) $N=1$, (b) $N=3$, (c) $N=5$, and (d) $N=7$. Red dashed circles mark the robust high-fidelity radius, whose expansion with increasing $N$ visualizes the enlarged error tolerance and reduced system sensitivity. The values of the error $\varepsilon$ are in units of $10^{-5}$.}
    \label{fig:kaxi_error}
\end{figure}

Similar to the previous section, we scan the fidelity variation of the target state as detuning errors are applied to both the cavity frequency $\omega_c' = \omega_c(1+\varepsilon)$ and the mechanical frequency $\omega_m' = \omega_m(1+\varepsilon)$. The results are presented in Fig.~\ref{fig:kaxi_error}, with each panel corresponding to a different number of parameter segments ($N = 1, 3, 5, 7$).

Consistent with the findings for three-photon resonance, the high-fidelity region for the benchmark scheme (Fig.~\ref{fig:kaxi_error}(a)) is confined to a vanishingly thin diagonal line. In contrast, our approach significantly expands this region, as visualized by the red dashed circles. The radius of these circles enlarges progressively with sequence depth, increasing from $\sim2.2 \times 10^{-8}$ to $\sim1 \times 10^{-7}$ for the $N=7$ case. The success achieved on such an extreme sensitivity scale (${\sim}10^{-7}$) demonstrates that the OPSS strategy could effectively reduce the sensitivity to detuning errors in both Hamiltonian parameters.

\begin{figure*}[tbp]
    \centering
    \begin{minipage}[b]{0.49\textwidth}
        \centering
        \includegraphics[width=\linewidth]{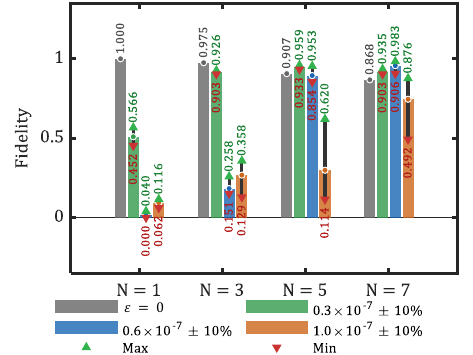} 
    \end{minipage}
    \hfill
    \begin{minipage}[b]{0.49\textwidth}
        \centering
        \includegraphics[width=\linewidth]{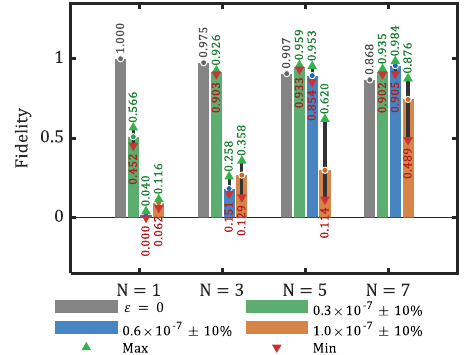} 
    \end{minipage}
    \caption{Performance and robustness analysis. (a) Statistical performance analysis of $\varepsilon$ on $\omega_m$ (with $\varepsilon$ on $\omega_c=0$). (b) Statistical robustness analysis of $\varepsilon$ on $\omega_c$ (with $\varepsilon$ on $\omega_m=0$). The height of each bar represents the mean fidelity, while the green upward triangles ($\blacktriangle$) and red downward triangles ($\blacktriangledown$) indicate the maximum and minimum fidelities within the defined ranges.}
    \label{fig:casimir_stats}
\end{figure*}

To verify the universality of our method, we performed a similar statistical robustness analysis for the Casimir-Rabi model. The error sensitivity here operates on a much finer scale (${\sim}10^{-7}$). We evaluated performance around three distinct error magnitudes: $0.3\times 10^{-7}$, $0.6\times 10^{-7}$, and $1.0 \times 10^{-7}$.

The results in Fig.~\ref{fig:casimir_stats} are similar to the trends observed in the three-photon resonance. The benchmark $N=1$ pulse is extremely fragile; even a minor error of $\varepsilon = 0.3 \times 10^{-7}$ causes the mean fidelity to drop to ${\sim}0.45$. Increasing the segment number $N$ effectively broadens the tolerance window. The $N=5$ sequence maintains robust performance ($F > 0.85$) up to $\varepsilon = 0.6 \times 10^{-7}$ but eventually degrades at larger magnitudes. The $N=7$ sequence delivers the most comprehensive protection. Notably, at the error magnitude of $\varepsilon = 1.0 \times 10^{-7}$, the maximum fidelity remains remarkably high ($> 0.87$) for both parameters. This confirms that the OPSS strategy successfully preserves high-fidelity solutions even in highly sensitive quantum systems.

\subsection{Output Photon Flux}
\label{sec:master_equation}
In this section, we employ the master equation to investigate the output photon flux of the cavity. We consider the three-photon resonance as an example, a system in which the artificial atom and the cavity field are coupled to separate baths. It is assumed that the temperature of the baths is zero. By employing the Born-Markov approximation, the master equation~\cite{Huang2014,Stassi2013,Ridolfo2012} for the reduced density matrix of the system, $\rho(t)$, in the Schr\"{o}dinger picture can be written as
\begin{equation}
    \frac{d\rho(t)}{dt} = -i[H(t), \rho(t)] + \kappa\mathcal{D}[X_1]\rho + \gamma\mathcal{D}[X_2]\rho.
    \label{eq:master_equation}
\end{equation}
Here, $H(t)$ is the total system Hamiltonian given by Eq.~\eqref{eq:Rabi_Hamiltonian}, $\kappa$ represents the photon decay rate, and $\gamma$ denotes the atomic relaxation rate. The operators $X_1$ and $X_2$ are defined by 
\begin{align}
    \label{eq:X1}
    X_1 &= \sum_{E_n > E_m} \langle \psi_m | (a + a^\dagger) | \psi_n \rangle \ket{\psi_m}\bra{\psi_n}, \\[5pt]
    \label{eq:X2_three}
    X_2 &= \sum_{E_n > E_m} \langle \psi_m | \sigma_x | \psi_n \rangle \ket{\psi_m}\bra{\psi_n}. 
\end{align}
$X_1$ and $X_2$ are the jump operators, describing transitions from high-energy eigenstates to low-energy eigenstates. The eigenstates $|\psi_n\rangle$ are the instantaneous eigenstates of the Hamiltonian, satisfying $H(t)\ket{\psi_n} = E_n(t)\ket{\psi_n}$. The standard Lindblad superoperator $\mathcal{D}$ is defined by
\begin{equation}
    \mathcal{D}[O]\rho = \frac{1}{2} \left( 2O\rho O^\dagger - O^\dagger O\rho - \rho O^\dagger O \right).
    \label{eq:lindblad_def}
\end{equation}

For the three-photon Rabi resonance, the output photon flux rate~\cite{Stassi2013,Ridolfo2012} from the cavity field is defined by
\begin{equation}
    \Phi_{\text{out}}(t) = \kappa \text{Tr}[\rho(t)X_1^\dagger X_1]. 
    \label{eq:flux_def}
\end{equation}

\begin{figure}[tbp]
    \centering
    \includegraphics[width=\columnwidth]{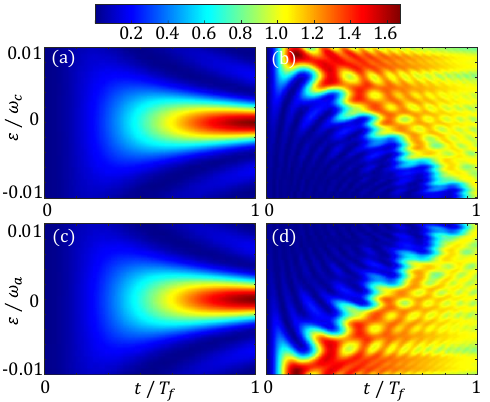}
    \caption{Differential output photon flux landscapes for the three-photon resonance model. The decay rates are set to $\kappa = \gamma = 0.00006$. The flux values are scaled by $10^{-4}$. (a) and (b) $\varepsilon$ on $\omega_a$. (c) and (d) $\varepsilon$ on $\omega_c$.}
    \label{photonflux_threephoton}
\end{figure}

To strictly evaluate the robustness of the scheme under realistic noise conditions, we calculate the time-dependent output photon flux using the global master equation approach. By mapping the output photon flux $\Phi_{\text{out}}$ against the detuning error $\varepsilon$, we demonstrate the dynamical behaviors of the three-photon resonance in Fig.~\ref{photonflux_threephoton}. We perform a comparative analysis of the output photon flux dynamics for the standard ($N=1$) and optimized ($N=7$) schemes in the presence of detuning errors. For clarity, the time evolution is plotted against the normalized ratio $t/T$, where $T$ denotes the total pulse duration. The duration for the unoptimized single-pulse case is $T_{N=1} \approx 1320\,\omega_c^{-1}$, whereas the OPSS requires a much longer duration of $T_{N=7} \approx 9440\,\omega_c^{-1}$.

As observed in the unoptimized three-photon resonance case in Figs.~\ref{photonflux_threephoton}(a) and \ref{photonflux_threephoton}(c), regardless of whether the error $\varepsilon$ is applied to $\omega_a$ or $\omega_c$, a significant output photon flux at the end of the evolution is confined to a narrow interval of $\varepsilon \in [-0.5\%, 0.5\%]$. This is consistent with our previous results shown in Fig.~\ref{fig:three_error}(a). In contrast, for the optimized $N=7$ case in Figs.~\ref{photonflux_threephoton}(b) and \ref{photonflux_threephoton}(d), a stable output photon flux is maintained at the end of the evolution across a much wider range of $\varepsilon \in [-1\%, 1\%]$. This demonstrates that the three-photon resonance can be maintained with our optimization strategy even in the presence of coupling to the environment. 

For the case of Casimir-Rabi resonance, the master equation remains identical to Eq.~\eqref{eq:master_equation}. The Hamiltonian is replaced by Eq.~\eqref{eq:optomech_hamiltonian}, where $\gamma$ represents the phonon decay rate. While $X_1$ remains unchanged, $X_2$ is defined as
\begin{equation}
    \label{eq:X2_Casimir}
    X_2 = \sum_{E_n > E_m} \langle \psi_m | (b + b^\dagger) | \psi_n \rangle \ket{\psi_m}\bra{\psi_n}.
\end{equation}

\section{EXPERIMENTAL IMPLEMENTATION}

To experimentally verify our theoretical predictions, we propose utilizing two distinct physical architectures tailored to the specific spectral and coupling requirements of the three-photon and Casimir-Rabi dynamics.

For the three-photon resonance, a superconducting circuit QED system is a promising candidate. We consider a fluxonium qubit coupled to a frequency-tunable transmission line resonator~\cite{Manucharyan2009,Chang2020}. The fluxonium circuit provides the necessary giant anharmonicity to resolve higher-order transitions, while the resonator frequency $\omega_c$ can be continuously tuned via a SQUID array termination~\cite{Sandberg2008}. This allows for the precise in-situ adjustment of the cavity mode to satisfy the three-photon resonance condition $\omega_q \approx 3\omega_c$. Typical experimental parameters for such setups involve qubit and cavity frequencies in the microwave domain ($\omega_{q,c}/2\pi \approx 4\text{--}8$~GHz) with tunable bandwidths exceeding $1$~GHz~\cite{Nguyen2019}, ensuring the system can be stabilized at the targeted triple-frequency optimization point.

The Casimir-Rabi oscillations can be effectively realized in a superconducting circuit architecture where a DC SQUID is integrated into a coplanar microwave cavity. In this setup, a segment of the SQUID loop functions as a mechanical resonator, establishing an effective optomechanical coupling via the flux-dependent Josephson inductance~\cite{FriskKockum2019, FornDaz2019}. By modulating the external magnetic flux, the effective cavity frequency is dynamically tuned to satisfy the resonance conditions necessary for the third-order phonon-photon interaction. Furthermore, considering the ultra-high quality factors of state-of-the-art nanomechanical resonators ($Q > 10^{10}$)~\cite{Dania2024} and superconducting cavities~\cite{Pietikainen2024}, this scheme is well within the reach of current experimental capabilities.

\section{CONCLUSION}
\label{sec:conclusion}

In conclusion, we have introduced and numerically validated a powerful optimization strategy to address the pronounced parameter sensitivity inherent in quantum systems governed by high-order interactions. By using OPSS, our method directly reduces the sensitivity of the three-photon resonance and the Casimir-Rabi resonance against detuning errors, expanding the high-fidelity operational window by an order of magnitude.

The consistent success across two systems with disparate physical mechanisms and sensitivity scales demonstrates the versatility and capability of our numerical discovery method. In contrast to traditional techniques that rely on simplified models~\cite{Ivanov2022} and analytical constructions, our strategy offers a more flexible and powerful paradigm for designing robust controls for complex quantum dynamics processes. This paves the way for harnessing high-order resonant phenomena for applications in quantum information processing and precision measurement in realistic, noisy environments.

Looking forward, this optimization framework can be extended in several promising directions. An immediate next step is to incorporate decoherence effects explicitly into the cost function, aiming to identify an optimal trade-off between error robustness and finite coherence times in open quantum systems~\cite{Ivanov2022, Koch2022}. Moreover, the methodology is readily applicable to other physical platforms grappling with similar parameter uncertainties, such as optimizing quantum gate operations in solid-state spin systems~\cite{Dolde2014} or trapped ions~\cite{Nebendahl2009}. We believe the numerical blueprints provided in this study will motivate future experimental efforts to achieve ultra-precise and robust control of high-order quantum processes.

\begin{acknowledgments}
Y.-H.C. was supported by the National Natural Science Foundation of China (Grant Nos.~12304390 and 12574386), the Fujian 100 Talents Program, and the Fujian Minjiang Scholar Program. Z.-C.S. was supported by the National Natural Science Foundation of China (Grant No.~62571129) and the Natural Science Foundation of Fujian Province (Grant No.~2025J01456). Y.X. was supported by the National Natural Science Foundation of China (Grant Nos.~11575045 and 62471143), the Natural Science Funds for Distinguished Young Scholar of Fujian Province (Grant No.~2020J06011), and the Project from Fuzhou University (Grant No.~JG202001-2).
\end{acknowledgments}

\section*{DATA AVAILABILITY}

The data used for obtaining the presented numerical results as well as for generating the
plots is available on request. Please refer to yehong.chen@fzu.edu.cn.

\appendix % 这一行表示从这里开始是附录

\section{DERIVATION OF THE EFFECTIVE HAMILTONIAN }
\label{app.A}
\subsection{THE THREE-PHOTON RESONANCE}

We begin with the quantum Rabi model in the interaction picture with respect to the free Hamiltonian $\hat{H}_0 = \frac{1}{2}\omega_a\hat{\sigma}_z + \omega_c\hat{a}^\dagger\hat{a}$. The interaction Hamiltonian ($\hbar = 1$) is given by
\begin{align}
    \hat{H}_I = \lambda[\hat{a}e^{i(\omega_a - \omega_c)t} + \hat{a}^\dagger e^{i(\omega_a + \omega_c)t}]\hat{\sigma}_+ + \text{H.c.},
    \label{eq:H_I_initial}
\end{align}
where $\lambda$ is the atom-cavity coupling rate, $\hat{a}$ ($\hat{a}^\dagger$) is the annihilation (creation) operator for the cavity mode of frequency $\omega_c$, and $\omega_a$ is the atomic transition frequency. The Pauli operators are defined as $\hat{\sigma}_z = |e\rangle\langle e| - |g\rangle\langle g|$, $\hat{\sigma}_+ = |e\rangle\langle g|$, and $\hat{\sigma}_- = |g\rangle\langle e|$.

Under the three-photon resonance condition $\omega_a = 3\omega_c$, the interaction Hamiltonian in Eq.~\eqref{eq:H_I_initial} can be expressed in terms of its harmonic components:
\begin{align}
    \hat{H}_I = \lambda(\hat{a}e^{2i\omega_c t} + \hat{a}^\dagger e^{4i\omega_c t}) \hat{\sigma}_+ + \text{H.c.}
    \label{eq:H_I_resonant}
\end{align}
From this expression, we can derive the effective Hamiltonian using the time-averaging method. According to this method, the second-order time-independent effective Hamiltonian is given by the general formula:
\begin{align}
    \hat{H}_{\text{eff}}^{(2)} = \sum_{m} \frac{1}{\omega_m} [\hat{h}_m, \hat{h}_m^\dagger],
    \label{eq:Heff2_general}
\end{align}
where $\hat{h}_m$ are the operator components of $\hat{H}_I$ oscillating at frequencies $\omega_m$. For the Hamiltonian in Eq.~\eqref{eq:H_I_resonant}, we can identify its harmonic components and, by substituting them into Eq.~\eqref{eq:Heff2_general}, obtain the second-order effective Hamiltonian:
\begin{align}
    \hat{H}_{\text{eff}}^{(2)} &= \frac{\lambda^2}{4\omega_c}[-(3\hat{a}^\dagger\hat{a} + 2)\hat{\sigma}_+\hat{\sigma}_- - (3\hat{a}^\dagger\hat{a} + 1)\hat{\sigma}_-\hat{\sigma}_+]. \label{eq:Heff_op_2}
\end{align}
Similarly, following the standard formalism~\cite{Huang2014, Shao2017}, the general formula for the third-order term is given by
\begin{equation}
\label{eq:Heff3_general}
\begin{split}
    \hat{H}_{\text{eff}}^{(3)}(t) = \sum_{l,m,n} \bigg\{
        & \frac{1}{\omega_n(\omega_n - \omega_m)} \Big[ \hat{h}_l \hat{h}_m^\dagger \hat{h}_n e^{i(\omega_l - \omega_m + \omega_n)t} \\
        & + \hat{h}_l^\dagger \hat{h}_m \hat{h}_n e^{i(-\omega_l + \omega_m + \omega_n)t} \\
        & + \hat{h}_l \hat{h}_m \hat{h}_n^\dagger e^{i(\omega_l + \omega_m - \omega_n)t} \\
        & + \hat{h}_l^\dagger \hat{h}_m^\dagger \hat{h}_n e^{i(-\omega_l - \omega_m + \omega_n)t} \Big] \\
        + & \frac{1}{\omega_n(\omega_n + \omega_m)} \Big[ \hat{h}_l \hat{h}_m^\dagger \hat{h}_n^\dagger e^{i(\omega_l - \omega_m - \omega_n)t} \\
        & + \hat{h}_l^\dagger \hat{h}_m \hat{h}_n^\dagger e^{i(-\omega_l + \omega_m - \omega_n)t} \Big]
    \bigg\}.
\end{split}
\end{equation}
which leads to the following time-averaged third-order correction:
\begin{align}
    \hat{H}_{\text{eff}}^{(3)} &= -\frac{\lambda^3}{4\omega_c^2}[(\hat{a}^\dagger)^3\hat{\sigma}_- + \hat{a}^3\hat{\sigma}_+]. \label{eq:Heff_op_3}
\end{align}

To obtain a concrete matrix representation, we project these operators onto the resonant two-level subspace spanned by the states $\{\ket{e, 0}, \ket{g, 3}\}$.
First, we evaluate the matrix elements of $\hat{H}_{\text{eff}}^{(2)}$. This operator is diagonal in this basis. The top-left element is $\bra{e, 0}\hat{H}_{\text{eff}}^{(2)}\ket{e, 0}$ ($n=0$):
\begin{align}
    (\hat{H}_{\text{eff}}^{(2)})_{11} = \frac{\lambda^2}{4\omega_c}[-(3(0) + 2)] = -\frac{\lambda^2}{2\omega_c}.
\end{align}
The bottom-right element is $\bra{g, 3}\hat{H}_{\text{eff}}^{(2)}\ket{g, 3}$ ($n=3$):
\begin{align}
    (\hat{H}_{\text{eff}}^{(2)})_{22} = \frac{\lambda^2}{4\omega_c}[-(3(3) + 1)] = -\frac{5\lambda^2}{2\omega_c}.
\end{align}

Next, we evaluate the matrix elements of $\hat{H}_{\text{eff}}^{(3)}$, which provides the off-diagonal coupling. The off-diagonal term is given by $\bra{e, 0}\hat{H}_{\text{eff}}^{(3)}\ket{g, 3}$:
\begin{align}
    (\hat{H}_{\text{eff}}^{(3)})_{12} &= \bra{e, 0} \left(-\frac{\lambda^3}{4\omega_c^2}\right) \hat{a}^3\hat{\sigma}_+ \ket{g, 3} \nonumber \\
    &= -\frac{\lambda^3}{4\omega_c^2} \bra{e, 0} \hat{a}^3 \ket{e, 3} \nonumber \\
    &= -\frac{\lambda^3}{4\omega_c^2} \bra{e, 0} \sqrt{3 \cdot 2 \cdot 1} \ket{e, 0} = -\frac{\sqrt{6}\lambda^3}{4\omega_c^2}.
\end{align}
The other off-diagonal element $(\hat{H}_{\text{eff}}^{(3)})_{21}$ is its Hermitian conjugate and has the same value.

Finally, the total effective Hamiltonian is the sum $\hat{H}_{\text{eff}} = \hat{H}_{\text{eff}}^{(2)} + \hat{H}_{\text{eff}}^{(3)}$. Its matrix representation in the basis $\{\ket{e, 0}, \ket{g, 3}\}$ is
\begin{align}
    \hat{H}_{\text{eff}} = 
    \begingroup
    \renewcommand{\arraystretch}{1.5} % 增加矩阵行高，使分式不拥挤
    \begin{pmatrix}
    -\dfrac{\lambda^2}{2\omega_c} & -\dfrac{\sqrt{6}\lambda^3}{4\omega_c^2} \\
    -\dfrac{\sqrt{6}\lambda^3}{4\omega_c^2} & -\dfrac{5\lambda^2}{2\omega_c}
    \end{pmatrix}. 
    \endgroup
    \label{eq:Heff_matrix_final}
\end{align}

The effective Hamiltonian in Eq.~\eqref{eq:Heff_matrix_final} is derived in the interaction picture, meaning its diagonal elements represent the energy shifts (AC Stark shifts) caused by the interaction. To obtain the total Hamiltonian in the laboratory frame (Schrödinger picture), we must add back the contribution from the free Hamiltonian, $\hat{H}_0$.
\begin{align}
    \hat{H}_{S} = \hat{H}_0 + \hat{H}_{\text{eff}}.
\end{align}
We compute the matrix elements of $\hat{H}_0$ in our basis. Applying the resonance condition $\omega_a = 3\omega_c$, this simplifies to $\frac{\omega_a}{2}$. Thus, $\hat{H}_0$ contributes a constant energy offset to both states.

By adding this diagonal contribution to Eq.~\eqref{eq:Heff_matrix_final}, we arrive at the final effective Hamiltonian in the Schrödinger picture:
\begin{align}
    \hat{H}_{S} = 
    \begingroup
    \renewcommand{\arraystretch}{1.5}
    \begin{pmatrix}
    \dfrac{\omega_a}{2} - \dfrac{\lambda^2}{2\omega_c} & - \dfrac{\sqrt{6}\lambda^3}{4\omega_c^2} \\
    - \dfrac{\sqrt{6}\lambda^3}{4\omega_c^2} & \dfrac{\omega_a}{2} - \dfrac{5\lambda^2}{2\omega_c}
    \end{pmatrix}. 
    \endgroup
    \label{eq:Heff_matrix_final_Schrodinger}
\end{align}

\subsection{The Casimir-Rabi Resonance}
\label{app.B}

To analyze the system dynamics, we transform the Hamiltonian into the interaction picture with respect to the free Hamiltonian $\hat{H}_0 = \omega_c \hat{a}^\dagger \hat{a} + \omega_m \hat{b}^\dagger \hat{b}$. In this picture, the interaction Hamiltonian $\hat{H}_I(t)$ can be expanded to explicitly show its harmonic components. Its full, unapproximated form is given by
\begin{align}
    \hat{H}_I(t) = g \Big[ &(\hat{a}^\dagger)^2 \hat{b}^\dagger e^{i(2\omega_c + \omega_m)t} + (\hat{a}^\dagger)^2 \hat{b} e^{i(2\omega_c - \omega_m)t} \nonumber \\
    & + (2\hat{a}^\dagger\hat{a} + 1)\hat{b}^\dagger e^{i\omega_m t} \Big] + \text{H.c.}
    \label{eq:casimir_H_I_compact}
\end{align}

According to Eq.~\eqref{eq:Heff2_general}, the second-order $\hat{H}_{\text{eff}}^{(2)}$ with the resonance condition $3\omega_m = 2\omega_c$ is given by
\begin{align}
    \hat{H}_{\text{eff}}^{(2)} = \frac{g^2}{\omega_m} \bigg[ &\frac{1}{4}\Big( (\hat{a}^\dagger)^2\hat{a}^2 \hat{b}^\dagger\hat{b} - \hat{b}\hat{b}^\dagger \hat{a}^2(\hat{a}^\dagger)^2 \Big) \nonumber \\
    + &\frac{1}{2}\Big( (\hat{a}^\dagger)^2\hat{a}^2 \hat{b}\hat{b}^\dagger - \hat{b}^\dagger\hat{b} \hat{a}^2(\hat{a}^\dagger)^2 \Big) \nonumber \\
    - & (2\hat{a}^\dagger\hat{a}+1)^2 \bigg].
    \label{eq:casimir_Heff2_operator_form}
\end{align}

Similarly, we derive the third-order effective Hamiltonian, $\hat{H}_{\text{eff}}^{(3)}$, by applying the general formula from Eq.~\eqref{eq:Heff3_general}. Instead of presenting its full and complex operator form, we focus on its matrix representation within the resonant subspace spanned by $\{\ket{2,0}, \ket{0,3}\}$. The evaluation of the off-diagonal matrix element $\bra{2,0} \hat{H}_{\text{eff}}^{(3)} \ket{0,3}$ yields the following compact matrix:
\begin{equation}
    \hat{H}_{\text{eff}}^{(3)}=\frac{18\sqrt{3}g^3}{\omega_m^2}
    \begin{pmatrix}
    0 & 1 \\
    1 & 0
    \end{pmatrix}.
    \label{eq:casimir_Heff3_matrix}
\end{equation}
Combining the second-order and the third-order terms, we obtain the full effective Hamiltonian in the interaction picture for the resonant subspace, $\hat{H}_{\text{eff}} = \hat{H}_{\text{eff}}^{(2)} + \hat{H}_{\text{eff}}^{(3)}$:
\begin{equation}
    \hat{H}_{\text{eff}} = 
    \begingroup
    \renewcommand{\arraystretch}{1.5}
    \begin{pmatrix}
    -\dfrac{27g^2}{\omega_m} & \dfrac{18\sqrt{3}g^3}{\omega_m^2} \\
    \dfrac{18\sqrt{3}g^3}{\omega_m^2} & -\dfrac{6g^2}{\omega_m}
    \end{pmatrix}.
    \endgroup
\end{equation}
To obtain the total Hamiltonian in the Schrödinger picture, we must add back the contribution from the free Hamiltonian, $\hat{H}_0 = \omega_c \hat{a}^\dagger \hat{a} + \omega_m \hat{b}^\dagger \hat{b}$. Its matrix elements in the basis $\{\ket{2,0}, \ket{0,3}\}$ are given by $E_1 = \bra{2,0}\hat{H}_0\ket{2,0} = 2\omega_c$ and $E_2 = \bra{0,3}\hat{H}_0\ket{0,3} = 3\omega_m$. 

The final effective Hamiltonian in the Schrödinger picture, $\hat{H}_{S} = \hat{H}_0 + \hat{H}_{\text{eff}}$, is therefore:
\begin{equation}
    \hat{H}_{S}= 
    \begingroup
    \renewcommand{\arraystretch}{1.5}
    \begin{pmatrix}
    2\omega_c - \dfrac{27g^2}{\omega_m} & \dfrac{18\sqrt{3}g^3}{\omega_m^2} \\
    \dfrac{18\sqrt{3}g^3}{\omega_m^2} & 3\omega_m - \dfrac{6g^2}{\omega_m}
    \end{pmatrix}.
    \endgroup
    \label{eq:casimir_Heff_final_Schrodinger}
\end{equation}
The diagonal elements of this matrix represent the absolute energies of the dressed states, while the off-diagonal terms describe the effective coupling between them.

\bibliography{reference}  

@article{FriskKockum2019,
  author  = {Frisk Kockum, Anton and Miranowicz, Adam and De Liberato, Simone and Savasta, Salvatore and Nori, Franco},
  title   = {Ultrastrong coupling between light and matter},
  journal = {Nat. Rev. Phys.},
  volume  = {1},
  pages   = {19--40},
  year    = {2019},
  doi     = {10.1038/s42254-018-0006-2}
}

@book{Lindsay1975,
  author    = {P. A. Lindsay},
  title     = {Introduction to Quantum Electronics},
  publisher = {Pitman},
  address   = {London},
  year      = {1975}
}

@book{Kielich1981,
  author    = {S. Kielich},
  title     = {Molecular Nonlinear Optics},
  publisher = {Nauka},
  address   = {Moscow},
  year      = {1981}
}

@book{Shen1984,
  author    = {Y. R. Shen},
  title     = {The Principles of Nonlinear Optics},
  publisher = {Wiley},
  address   = {New York},
  year      = {1984}
}

@book{Boyd2008,
  author    = {Robert W. Boyd},
  title     = {Nonlinear Optics},
  edition   = {3rd},
  publisher = {Elsevier},
  address   = {Amsterdam},
  year      = {2008}
}

@article{Liu2022,
  author  = {Liu, Lijuan and Zhao, Lin and Zhou, Xingjiang and Wang, Xiaoyang},
  title   = {Recent progress in the development of {KBe2BO3F2}: a deep-UV nonlinear optical crystal},
  journal = {Appl. Phys. B},
  volume  = {128},
  pages   = {17},
  year    = {2022},
  doi     = {10.1007/s00340-021-07744-0}
}

@article{Dudley2006,
  author  = {Dudley, John M. and Genty, Go\"{e}ry and Coen, St\'{e}phane},
  title   = {Supercontinuum generation in photonic crystal fiber},
  journal = {Rev. Mod. Phys.},
  volume  = {78},
  pages   = {1135--1184},
  year    = {2006},
  doi     = {10.1103/revmodphys.78.1135}
}

@article{Acharyya2021,
  author  = {Acharyya, Jitendra Nath and Prakash, G. Vijaya},
  title   = {Nonlinear optical dispersion and higher-order effects in bulk and wavelength-ordered photonic materials},
  journal = {Optik},
  volume  = {247},
  pages   = {167944},
  year    = {2021},
  doi     = {10.1016/j.ijleo.2021.167944}
}

@article{Kumar2021,
  author  = {Kumar, Vipin},
  title   = {Linear and Nonlinear Optical Properties of Graphene: A Review},
  journal = {J. Electron. Mater.},
  volume  = {50},
  pages   = {3773--3799},
  year    = {2021},
  doi     = {10.1007/s11664-021-08904-w}
}

@book{2024,
  editor    = {Hamilton, Mark F. and Blackstock, David T.},
  title     = {Nonlinear Acoustics},
  publisher = {Springer Nature Switzerland},
  address   = {Cham},
  year      = {2024},
  doi       = {}
}

@article{Rolston2002,
  author  = {Rolston, S. L. and Phillips, W. D.},
  title   = {Nonlinear and quantum atom optics},
  journal = {Nature},
  volume  = {416},
  pages   = {219--224},
  year    = {2002},
  doi     = {10.1038/416219a}
}

@article{Savelev2006,
  author  = {Savel'ev, Sergey and Rakhmanov, A. L. and Yampol'skii, V. A. and Nori, Franco},
  title   = {Analogues of nonlinear optics using terahertz {Josephson} plasma waves in layered superconductors},
  journal = {Nat. Phys.},
  volume  = {2},
  pages   = {521--525},
  year    = {2006},
  doi     = {10.1038/nphys358}
}

@article{Blais2021,
  author  = {Blais, Alexandre and Grimsmo, Arne L. and Girvin, S. M. and Wallraff, Andreas},
  title   = {Circuit quantum electrodynamics},
  journal = {Rev. Mod. Phys.},
  volume  = {93},
  pages   = {025005},
  year    = {2021},
  doi     = {10.1103/RevModPhys.93.025005}
}

@article{Niemczyk2010,
  author  = {Niemczyk, T. and Deppe, F. and Huebl, H. and Menzel, E. P. and Hocke, F. and Schwarz, M. J. and Garcia-Ripoll, J. J. and Zueco, D. and H\"{u}mmer, T. and Solano, E. and Marx, A. and Gross, R.},
  title   = {Circuit quantum electrodynamics in the ultrastrong-coupling regime},
  journal = {Nat. Phys.},
  volume  = {6},
  pages   = {772--776},
  year    = {2010},
  doi     = {10.1038/nphys1730}
}

@article{Fedorov2010,
  author  = {Fedorov, A. and Feofanov, A. K. and Macha, P. and Forn-D\'{\i}az, P. and Harmans, C. J. P. M. and Mooij, J. E.},
  title   = {Strong Coupling of a Quantum Oscillator to a Flux Qubit at Its Symmetry Point},
  journal = {Phys. Rev. Lett.},
  volume  = {105},
  pages   = {060503},
  year    = {2010},
  doi     = {10.1103/PhysRevLett.105.060503}
}

@article{FornDaz2016,
  author  = {Forn-D\'{\i}az, P. and Romero, G. and Harmans, C. J. P. M. and Solano, E. and Mooij, J. E.},
  title   = {Broken selection rule in the quantum {Rabi} model},
  journal = {Sci. Rep.},
  volume  = {6},
  pages   = {26720},
  year    = {2016},
  doi     = {10.1038/srep26720}
}

@article{FornDaz2019,
  author  = {Forn-D\'{\i}az, P. and Lamata, L. and Rico, E. and Kono, J. and Solano, E.},
  title   = {Ultrastrong coupling regimes of light-matter interaction},
  journal = {Rev. Mod. Phys.},
  volume  = {91},
  pages   = {025005},
  year    = {2019},
  doi     = {10.1103/RevModPhys.91.025005}
}

@article{FornDaz2010,
  author  = {Forn-D\'{\i}az, P. and Lisenfeld, J. and Marcos, D. and Garc\'{\i}a-Ripoll, J. J. and Solano, E. and Harmans, C. J. P. M. and Mooij, J. E.},
  title   = {Observation of the {Bloch}-{Siegert} Shift in a Qubit-Oscillator System in the Ultrastrong Coupling Regime},
  journal = {Phys. Rev. Lett.},
  volume  = {105},
  pages   = {237001},
  year    = {2010},
  doi     = {10.1103/PhysRevLett.105.237001}
}

@article{Garziano2016,
  author  = {Garziano, Luigi and Macrì, Vincenzo and Stassi, Roberto and Di Stefano, Omar and Nori, Franco and Savasta, Salvatore},
  title   = {One Photon Can Simultaneously Excite Two or More Atoms},
  journal = {Phys. Rev. Lett.},
  volume  = {117},
  pages   = {043601},
  year    = {2016},
  doi     = {10.1103/PhysRevLett.117.043601}
}

@article{Garziano2015,
  author  = {Garziano, Luigi and Stassi, Roberto and Macrì, Vincenzo and Kockum, Anton Frisk and Savasta, Salvatore and Nori, Franco},
  title   = {Multiphoton quantum {Rabi} oscillations in ultrastrong cavity {QED}},
  journal = {Phys. Rev. A},
  volume  = {92},
  pages   = {063830},
  year    = {2015},
  doi     = {10.1103/PhysRevA.92.063830}
}

@article{Ma2015,
  author  = {Ma, Ken K. W. and Law, C. K.},
  title   = {Three-photon resonance and adiabatic passage in the large-detuning {Rabi} model},
  journal = {Phys. Rev. A},
  volume  = {92},
  pages   = {023842},
  year    = {2015},
  doi     = {10.1103/PhysRevA.92.023842}
}

@article{Yan2025,
  author  = {Yan, Ke-Xiong and Qiu, Yuan and Xiao, Yang and Song, Jie and Chen, Ye-Hong and Xia, Yan},
  title   = {Spontaneous emission in {Casimir}-{Rabi} oscillations through a weak optomechanical coupling},
  journal = {Opt. Express},
  volume  = {33},
  pages   = {39283},
  year    = {2025},
  doi     = {10.1364/OE.563278}
}

@book{Freeman1998,
  author    = {Freeman, Ray},
  title     = {Spin Choreography: Basic Steps in High Resolution {NMR}},
  publisher = {Oxford University Press},
  address   = {Oxford},
  year      = {1998},
  doi       = {}
}

@article{Levitt2007,
  author  = {Levitt, Malcolm H.},
  title   = {Composite Pulses},
  journal = {Encycl. Magn. Reson.},
  pages   = {emrstm0086},
  year    = {2007},
  doi     = {10.1002/9780470034590.emrstm0086}
}

@article{HAFFNER2008,
  author  = {H\"affner, H. and Roos, C. F. and Blatt, R.},
  title   = {Quantum computing with trapped ions},
  journal = {Phys. Rep.},
  volume  = {469},
  pages   = {155--203},
  year    = {2008},
  doi     = {10.1016/j.physrep.2008.09.003}
}

@article{Gulde2003,
  author  = {Gulde, Stephan and Riebe, Mark and Lancaster, Gavin P. T. and Becher, Christoph and Eschner, J\"{u}rgen and H\"{a}ffner, Hartmut and Schmidt-Kaler, Ferdinand and Chuang, Isaac L. and Blatt, Rainer},
  title   = {Implementation of the {Deutsch}-{Jozsa} algorithm on an ion-trap quantum computer},
  journal = {Nature},
  volume  = {421},
  pages   = {48--50},
  year    = {2003},
  doi     = {10.1038/nature01336}
}

@article{Timoney2008,
  author  = {Timoney, N. and Elman, V. and Glaser, S. and Weiss, C. and Johanning, M. and Neuhauser, W. and Wunderlich, Chr.},
  title   = {Error-resistant single-qubit gates with trapped ions},
  journal = {Phys. Rev. A},
  volume  = {77},
  pages   = {052334},
  year    = {2008},
  doi     = {10.1103/PhysRevA.77.052334}
}

@article{Monz2009,
  author  = {Monz, T. and Kim, K. and H\"{a}nsel, W. and Riebe, M. and Villar, A. S. and Schindler, P. and Chwalla, M. and Hennrich, M. and Blatt, R.},
  title   = {Realization of the Quantum {Toffoli} Gate with Trapped Ions},
  journal = {Phys. Rev. Lett.},
  volume  = {102},
  pages   = {040501},
  year    = {2009},
  doi     = {10.1103/PhysRevLett.102.040501}
}

@article{Shappert2013,
  author  = {Shappert, C. M. and Merrill, J. T. and Brown, K. R. and Amini, J. M. and Volin, C. and Doret, S. C. and Hayden, H. and Pai, C.-S. and Brown, K. R. and Harter, A. W.},
  title   = {Spatially uniform single-qubit gate operations with near-field microwaves and composite pulse compensation},
  journal = {New J. Phys.},
  volume  = {15},
  pages   = {083053},
  year    = {2013},
  doi     = {10.1088/1367-2630/15/8/083053}
}

@article{Mount2015,
  author  = {Mount, Emily and Kabytayev, Chingiz and Crain, Stephen and Harper, Robin and Baek, So-Young and Vrijsen, Geert and Flammia, Steven T. and Brown, Kenneth R. and Maunz, Peter and Kim, Jungsang},
  title   = {Error compensation of single-qubit gates in a surface-electrode ion trap using composite pulses},
  journal = {Phys. Rev. A},
  volume  = {92},
  pages   = {060301},
  year    = {2015},
  doi     = {10.1103/PhysRevA.92.060301}
}

@article{Zarantonello2019,
  author  = {Zarantonello, G. and Hahn, H. and Morgner, J. and Schulte, M. and Bautista-Salvador, A. and Werner, R. F. and Hammerer, K. and Ospelkaus, C.},
  title   = {Robust and Resource-Efficient Microwave Near-Field Entangling {${}^{9}\text{Be}^{+}$} Gate},
  journal = {Phys. Rev. Lett.},
  volume  = {123},
  pages   = {260503},
  year    = {2019},
  doi     = {10.1103/PhysRevLett.123.260503}
}

@article{Rakreungdet2009,
  author  = {Rakreungdet, Worawarong and Lee, Jae Hoon and Lee, Kim Fook and Mischuck, Brian E. and Montano, Enrique and Jessen, Poul S.},
  title   = {Accurate microwave control and real-time diagnostics of neutral-atom qubits},
  journal = {Phys. Rev. A},
  volume  = {79},
  pages   = {022316},
  year    = {2009},
  doi     = {10.1103/PhysRevA.79.022316}
}

@article{Demeter2016,
  author  = {Demeter, Gabor},
  title   = {Composite pulses for high-fidelity population inversion in optically dense, inhomogeneously broadened atomic ensembles},
  journal = {Phys. Rev. A},
  volume  = {93},
  pages   = {023830},
  year    = {2016},
  doi     = {10.1103/PhysRevA.93.023830}
}

@article{Wang2012,
  author  = {Wang, Xin and Bishop, Lev S. and Kestner, J. P. and Barnes, Edwin and Sun, Kai and Das Sarma, S.},
  title   = {Composite pulses for robust universal control of singlet-triplet qubits},
  journal = {Nat. Commun.},
  volume  = {3},
  pages   = {997},
  year    = {2012},
  doi     = {10.1038/ncomms2003}
}

@article{Eng2015,
  author  = {Eng, Kevin and Ladd, Thaddeus D. and Smith, Aaron and Borselli, Matthew G. and Kiselev, Andrey A. and Fong, Bryan H. and Holabird, Kevin S. and Hazard, Thomas M. and Huang, Biqin and Deelman, Peter W. and Milosavljevic, Ivan and Schmitz, Adele E. and Ross, Richard S. and Gyure, Mark F. and Hunter, Andrew T.},
  title   = {Isotopically enhanced triple-quantum-dot qubit},
  journal = {Sci. Adv.},
  volume  = {1},
  pages   = {e1500214},
  year    = {2015},
  doi     = {10.1126/sciadv.1500214}
}

@article{Genov2017,
  author  = {Genov, Genko T. and Schraft, Daniel and Vitanov, Nikolay V. and Halfmann, Thomas},
  title   = {Arbitrarily Accurate Pulse Sequences for Robust Dynamical Decoupling},
  journal = {Phys. Rev. Lett.},
  volume  = {118},
  pages   = {133202},
  year    = {2017},
  doi     = {10.1103/PhysRevLett.118.133202}
}

@article{Wang2014,
  author  = {Wang, Xin and Bishop, Lev S. and Barnes, Edwin and Kestner, J. P. and Das Sarma, S.},
  title   = {Robust quantum gates for singlet-triplet spin qubits using composite pulses},
  journal = {Phys. Rev. A},
  volume  = {89},
  pages   = {022310},
  year    = {2014},
  doi     = {10.1103/PhysRevA.89.022310}
}

@article{Zhang2017,
  author  = {Zhang, Chengxian and Throckmorton, Robert E. and Yang, Xu-Chen and Wang, Xin and Barnes, Edwin and Das Sarma, S.},
  title   = {Randomized Benchmarking of Barrier versus Tilt Control of a Singlet-Triplet Qubit},
  journal = {Phys. Rev. Lett.},
  volume  = {118},
  pages   = {216802},
  year    = {2017},
  doi     = {10.1103/PhysRevLett.118.216802}
}

@article{Kestner2013,
  author  = {Kestner, J. P. and Wang, Xin and Bishop, Lev S. and Barnes, Edwin and Das Sarma, S.},
  title   = {Noise-Resistant Control for a Spin Qubit Array},
  journal = {Phys. Rev. Lett.},
  volume  = {110},
  pages   = {140502},
  year    = {2013},
  doi     = {10.1103/PhysRevLett.110.140502}
}

@article{Hickman2013,
  author  = {Hickman, G. T. and Wang, Xin and Kestner, J. P. and Das Sarma, S.},
  title   = {Dynamically corrected gates for an exchange-only qubit},
  journal = {Phys. Rev. B},
  volume  = {88},
  pages   = {161303},
  year    = {2013},
  doi     = {10.1103/PhysRevB.88.161303}
}

@article{Bilal2020,
  author  = {Bilal and Pant, Millie and Zaheer, Hira and Garcia-Hernandez, Laura and Abraham, Ajith},
  title   = {Differential {Evolution}: A review of more than two decades of research},
  journal = {Eng. Appl. Artif. Intell.},
  volume  = {90},
  pages   = {103479},
  year    = {2020},
  doi     = {10.1016/j.engappai.2020.103479}
}

@article{Ahmad2022,
  author  = {Ahmad, Mohamad Faiz and Isa, Nor Ashidi Mat and Lim, Wei Hong and Ang, Koon Meng},
  title   = {Differential {Evolution}: A recent review based on state-of-the-art works},
  journal = {Alexandria Eng. J.},
  volume  = {61},
  pages   = {3831--3872},
  year    = {2022},
  doi     = {10.1016/j.aej.2021.09.013}
}

@article{Khaneja2005,
  author  = {Khaneja, Navin and Reiss, Timo and Kehlet, Cindie and Schulte-Herbr\"{u}ggen, Thomas and Glaser, Steffen J.},
  title   = {Optimal control of coupled spin dynamics: design of {NMR} pulse sequences by gradient ascent algorithms},
  journal = {J. Magn. Reson.},
  volume  = {172},
  pages   = {296--305},
  year    = {2005},
  doi     = {10.1016/j.jmr.2004.11.004}
}

@article{Machnes2011,
  author  = {Machnes, S. and Sander, U. and Glaser, S. J. and de Fouquières, P. and Gruslys, A. and Schirmer, S. and Schulte-Herbr\"{u}ggen, T.},
  title   = {Comparing, optimizing, and benchmarking quantum-control algorithms in a unifying programming framework},
  journal = {Phys. Rev. A},
  volume  = {84},
  pages   = {022305},
  year    = {2011},
  doi     = {10.1103/PhysRevA.84.022305}
}

@article{Shao2017,
  author  = {Shao, Wenjun and Wu, Chunfeng and Feng, Xun-Li},
  title   = {Generalized {James}' effective {Hamiltonian} method},
  journal = {Phys. Rev. A},
  volume  = {95},
  pages   = {032124},
  year    = {2017},
  doi     = {10.1103/PhysRevA.95.032124}
}

@article{Law1995,
  author  = {Law, C. K.},
  title   = {Interaction between a moving mirror and radiation pressure: A {Hamiltonian} formulation},
  journal = {Phys. Rev. A},
  volume  = {51},
  pages   = {2537--2541},
  year    = {1995},
  doi     = {10.1103/PhysRevA.51.2537}
}

@article{Ivanov2022,
  author  = {Ivanov, Svetoslav S. and Torosov, Boyan T. and Vitanov, Nikolay V.},
  title   = {High-Fidelity Quantum Control by Polychromatic Pulse Trains},
  journal = {Phys. Rev. Lett.},
  volume  = {129},
  pages   = {240505},
  year    = {2022},
  doi     = {10.1103/PhysRevLett.129.240505}
}

@article{Byrd1995,
  author  = {Byrd, Richard H. and Lu, Peihuang and Nocedal, Jorge and Zhu, Ciyou},
  title   = {A Limited Memory Algorithm for Bound Constrained Optimization},
  journal = {SIAM J. Sci. Comput.},
  volume  = {16},
  pages   = {1190--1208},
  year    = {1995},
  doi     = {10.1137/0916069}
}

@article{Zhu1997,
  author  = {Zhu, Ciyou and Byrd, Richard H. and Lu, Peihuang and Nocedal, Jorge},
  title   = {Algorithm 778: {L-BFGS-B}: {Fortran} subroutines for large-scale bound-constrained optimization},
  journal = {ACM Trans. Math. Softw.},
  volume  = {23},
  pages   = {550--560},
  year    = {1997},
  doi     = {10.1145/279232.279236}
}

@misc{Mueller2025,
  author        = {Mueller, Niclas S. and Barros, Eduardo B. and Reich, Stephanie},
  title         = {Ultrastrong Light-Matter Coupling in Materials},
  year          = {2025},
  eprint        = {2505.06373},
  archivePrefix = {arXiv},
  primaryClass  = {physics.optics},
  doi           = {}
}

@article{Ridolfo2012,
  author  = {Ridolfo, A. and Leib, M. and Savasta, S. and Hartmann, M. J.},
  title   = {Photon Blockade in the Ultrastrong Coupling Regime},
  journal = {Phys. Rev. Lett.},
  volume  = {109},
  pages   = {193602},
  year    = {2012},
  doi     = {10.1103/PhysRevLett.109.193602}
}

@article{Stassi2013,
  author  = {Stassi, R. and Ridolfo, A. and Di Stefano, O. and Hartmann, M. J. and Savasta, S.},
  title   = {Spontaneous Conversion from Virtual to Real Photons in the Ultrastrong-Coupling Regime},
  journal = {Phys. Rev. Lett.},
  volume  = {110},
  pages   = {243601},
  year    = {2013},
  doi     = {10.1103/PhysRevLett.110.243601}
}

@article{Huang2014,
  author  = {Huang, Jin-Feng and Law, C. K.},
  title   = {Photon emission via vacuum-dressed intermediate states under ultrastrong coupling},
  journal = {Phys. Rev. A},
  volume  = {89},
  pages   = {033827},
  year    = {2014},
  doi     = {10.1103/PhysRevA.89.033827}
}

@article{Kyoseva2019,
  author  = {Kyoseva, Elica and Greener, Hadar and Suchowski, Haim},
  title   = {Detuning-modulated composite pulses for high-fidelity robust quantum control},
  journal = {Phys. Rev. A},
  volume  = {100},
  pages   = {032333},
  year    = {2019},
  doi     = {10.1103/PhysRevA.100.032333}
}

@article{Torosov2019,
  author  = {Torosov, Boyan T. and Vitanov, Nikolay V.},
  title   = {Robust high-fidelity coherent control of two-state systems by detuning pulses},
  journal = {Phys. Rev. A},
  volume  = {99},
  pages   = {013424},
  year    = {2019},
  doi     = {10.1103/PhysRevA.99.013424}
}

@article{Yan2024,
  author  = {Yan, Ke-Xiong and Qiu, Yuan and Xiao, Yang and Chen, Ye-Hong and Xia, Yan},
  title   = {Generating three-photon {Rabi} oscillations without a large-detuning condition},
  journal = {Phys. Rev. A},
  volume  = {110},
  pages   = {043711},
  year    = {2024},
  doi     = {10.1103/PhysRevA.110.043711}
}

@article{Macr2018,
  author  = {Macr\`{i}, Vincenzo and Ridolfo, Alessandro and Di Stefano, Omar and Kockum, Anton Frisk and Nori, Franco and Savasta, Salvatore},
  title   = {Nonperturbative Dynamical {Casimir} Effect in Optomechanical Systems: Vacuum {Casimir}-{Rabi} Splittings},
  journal = {Phys. Rev. X},
  volume  = {8},
  pages   = {011031},
  year    = {2018},
  doi     = {10.1103/PhysRevX.8.011031}
}

@article{Manucharyan2009,
  author  = {Manucharyan, Vladimir E. and Koch, Jens and Glazman, Leonid I. and Devoret, Michel H.},
  title   = {Fluxonium: Single Cooper-Pair Circuit Free of Charge Offsets},
  journal = {Science},
  volume  = {326},
  pages   = {113--116},
  year    = {2009},
  doi     = {10.1126/science.1175552}
}

@article{Chang2020,
  author  = {Chang, C. W. Sandbo and Sab\'{\i}n, Carlos and Forn-D\'{\i}az, P. and Quijandr\'{\i}a, Fernando and Vadiraj, A. M. and Nsanzineza, I. and Johansson, G. and Wilson, C. M.},
  title   = {Observation of Three-Photon Spontaneous Parametric Down-Conversion in a Superconducting Parametric Cavity},
  journal = {Phys. Rev. X},
  volume  = {10},
  pages   = {011011},
  year    = {2020},
  doi     = {10.1103/PhysRevX.10.011011}
}

@article{Sandberg2008,
  author  = {Sandberg, M. and Wilson, C. M. and Persson, F. and Bauch, T. and Johansson, G. and Shumeiko, V. and Duty, T. and Delsing, P.},
  title   = {Tuning the field in a microwave resonator faster than the photon lifetime},
  journal = {Appl. Phys. Lett.},
  volume  = {92},
  pages   = {203501},
  year    = {2008},
  doi     = {10.1063/1.2929367}
}

@article{Nguyen2019,
  author  = {Nguyen, Long B. and Lin, Yen-Hsiang and Somoroff, Aaron and Mencia, Raymond and Grabon, Nicholas and Manucharyan, Vladimir E.},
  title   = {High-Coherence Fluxonium Qubit},
  journal = {Phys. Rev. X},
  volume  = {9},
  pages   = {041041},
  year    = {2019},
  doi     = {10.1103/PhysRevX.9.041041}
}

@article{Torosov2011,
  author  = {Torosov, Boyan T. and Vitanov, Nikolay V.},
  title   = {Smooth composite pulses for high-fidelity quantum information processing},
  journal = {Phys. Rev. A},
  volume  = {83},
  pages   = {053420},
  year    = {2011},
  doi     = {10.1103/PhysRevA.83.053420}
}

@article{Das2011,
  author  = {Das, Swagatam and Suganthan, Ponnuthurai Nagaratnam},
  title   = {{Differential Evolution}: A Survey of the State-of-the-Art},
  journal = {IEEE Trans. Evol. Comput.},
  volume  = {15},
  pages   = {4--31},
  year    = {2011},
  doi     = {10.1109/tevc.2010.2059031}
}

@article{Storn1997,
  author  = {Storn, Rainer and Price, Kenneth},
  title   = {{Differential Evolution} -- A Simple and Efficient Heuristic for global Optimization over Continuous Spaces},
  journal = {J. Global Optim.},
  volume  = {11},
  pages   = {341--359},
  year    = {1997},
  doi     = {10.1023/a:1008202821328}
}

@article{Dania2024,
  author  = {Dania, Lorenzo and Bykov, Dmitry S. and Goschin, Florian and Teller, Markus and Kassid, Abderrahmane and Northup, Tracy E.},
  title   = {Ultrahigh Quality Factor of a Levitated Nanomechanical Oscillator},
  journal = {Phys. Rev. Lett.},
  volume  = {132},
  pages   = {133602},
  year    = {2024},
  doi     = {10.1103/PhysRevLett.132.133602}
}

@misc{Pietikainen2024,
  author        = {Pietik\"{a}inen, Iivari and \v{C}ernot\'{\i}k, Ond\v{r}ej and Eickbusch, Alec and Maiti, Aniket and Garmon, John W. O. and Filip, Radim and Girvin, Steven M.},
  title         = {Strategies and trade-offs for controllability and memory time of ultra-high-quality microwave cavities in circuit {QED}},
  year          = {2024},
  eprint        = {2403.02278},
  archivePrefix = {arXiv},
  primaryClass  = {quant-ph}
}

@article{Koch2022,
  author  = {Koch, Christiane P. and Boscain, Ugo and Calarco, Tommaso and Dirr, Gunther and Filipp, Stefan and Glaser, Steffen J. and Kosloff, Ronnie and Montangero, Simone and Schulte-Herbr\"{u}ggen, Thomas and Sugny, Dominique and Wilhelm, Frank K.},
  title   = {Quantum optimal control in quantum technologies. Strategic report on current status, visions and goals for research in {Europe}},
  journal = {EPJ Quantum Technol.},
  volume  = {9},
  pages   = {19},
  year    = {2022},
  doi     = {10.1140/epjqt/s40507-022-00138-x}
}

@article{Nebendahl2009,
  author  = {Nebendahl, V. and H\"{a}ffner, H. and Roos, C. F.},
  title   = {Optimal control of entangling operations for trapped-ion quantum computing},
  journal = {Phys. Rev. A},
  volume  = {79},
  pages   = {012312},
  year    = {2009},
  doi     = {10.1103/PhysRevA.79.012312}
}

@article{Dolde2014,
  author  = {Dolde, Florian and Bergholm, Ville and Wang, Ya and Jakobi, Ingmar and Naydenov, Boris and Pezzagna, S\'{e}bastien and Meijer, Jan and Jelezko, Fedor and Neumann, Philipp and Schulte-Herbr\"{u}ggen, Thomas and Biamonte, Jacob and Wrachtrup, J\"{o}rg},
  title   = {High-fidelity spin entanglement using optimal control},
  journal = {Nat. Commun.},
  volume  = {5},
  pages   = {3371},
  year    = {2014},
  doi     = {10.1038/ncomms4371}
}
\end{document}